# Superconductivity at 7.8 K in the ternary $LaRu_2As_2$ compound

Qi Guo[1], Bo-Jin Pan[1], Jia Yu[1], Bin-Bin Ruan[1], Dong-Yun Chen[1], Xiao-Chuan Wang[1], Qing-Ge Mu[1], Gen-Fu Chen[1,2], Zhi-An Ren[1,2],[*]

[1] Institute of Physics and Beijing National Laboratory for Condensed Matter Physics, Chinese Academy of Sciences, Beijing 100190, China

[2] Collaborative Innovation Center of Quantum Matter, Beijing 100190, China

[*] Email: renzhian@iphy.ac.cn

**Abstract**

Here we report the discovery of superconductivity in the ternary $LaRu_2As_2$ compound. The polycrystalline $LaRu_2As_2$ samples were synthesized by the conventional solid state reaction method. Powder X-ray diffraction analysis indicates that $LaRu_2As_2$ crystallizes in the $ThCr_2Si_2$-type crystal structure with the space group $I4/mmm$ (No. 139), and the refined lattice parameters are $a$ = 4.182(6) Å and $c$ = 10.590(3) Å. The temperature dependent resistivity measurement shows a clear superconducting transition with the onset $T_c$ (critical temperature) at 7.8 K, and zero resistivity happens at 6.8 K. The upper critical field at zero temperature $\mu_0 H_{c2}(0)$ was estimated to be 1.6 T from the resistivity measurement. DC magnetic susceptibility measurement shows a bulk superconducting Meissner transition at 7.0 K, and the isothermal magnetization measurement indicates that $LaRu_2As_2$ is a type-II superconductor.

The ternary compounds with body-centered tetragonal $ThCr_2Si_2$-type layered crystal structure consist of hundreds of different materials with rich physical properties, and most of them are based on transition metal elements with some of them being superconductors at low temperature [1, 2]. Recently high temperature superconductivity was discovered in iron arsenides upon the suppression of long-range antiferromagnetic ordering in several types of undoped parent compounds such as the quaternary LaFeAsO by ways of carrier doping *etc.* [3-5]. In these iron-based materials the intermetallic ternary 122-type compounds $AFe_2As_2$ (A = Ca, Sr, Ba, *etc.*) belong to the $ThCr_2Si_2$-type crystal structure and are free of oxygen with metallic behavior [5-7], and these superconductors have been widely studied for the investigation of their high-$T_c$ superconducting mechanism [8, 9]. Since the element Ru locates in the same group with Fe in the periodic table, similar superconductivity was expected in the Ru-based compounds, and some studies have been focused on the Ru-related compounds.

For the Ru-based superconductors, the layered perovskite ruthenium oxide $Sr_2RuO_4$ is a well-known realization for the spin-triplet p-wave pairing superconductor with a $T_c$ of 0.93 K [10]. The ternary compound $LaRu_2P_2$ superconductor crystallizes in the $ThCr_2Si_2$-type structure and has a $T_c$ of 4.1 K [2]. Superconductivity was also found in skutterudite compounds of $LaRu_4P_{12}$ and $LaRu_4As_{12}$ with $T_c$ at 7.2 K and 10.3 K [11, 12]. Superconductivity of higher $T_c$ at 13 K and 12 K was found in ZrRuP and ZrRuAs with $Fe_2P$-type structure respectively [13, 14]. Metal-insulator transition was reported in the binary compounds RuP and RuAs with MnP-type structure, and it was suppressed by Rh-doping at the Ru-site, which also led to superconductivity in them [15]. The 122-type $BaRu_2As_2$ and $SrRu_2As_2$ compounds with $ThCr_2Si_2$-type structure were found to be metallic, non-superconducting and show no signature of long-range magnetic ordering with measurements down to 1.8 K [16]. The 1111-type LnRuAsO (Ln = rare earth elements) compounds were also carefully investigated for structural and physical properties, and antiferromagnetic to ferromagnetic order transition was found in some of these compounds, while no superconductivity was ever reported, even in the F-doped materials [17-21]. Besides, for other iron-free pnictides with

ThCr$_2$Si$_2$-type crystal structure, superconductivity was also reported in SrNi$_2$As$_2$, BaNi$_2$As$_2$, SrIr$_2$As$_2$, CaPd$_2$As$_2$, BaPd$_2$As$_2$, LaPd$_2$As$_2$, *etc*., while their $T_c$ is all lower than 5 K [22-27].

We recently studied the Ru-based compounds, and synthesized the polycrystalline samples of ThCr$_2$Si$_2$-type LaRu$_2$As$_2$. Here we report the bulk superconductivity in LaRu$_2$As$_2$ revealed by resistivity and magnetization characterizations with an onset $T_c$ of 7.8 K.

The polycrystalline LaRu$_2$As$_2$ samples were synthesized by a conventional solid state reaction method. The starting materials of La pieces, Ru and As powders were mixed together in the stoichiometric ratio, pressed into pellets, and sealed in evacuated quartz tubes (or Ar-filled tantalum tubes). The sintering temperature was between 900 °C and 1100 °C, with a sintering period of 100 hours. The sintering procedures were repeated for at least 3 times for a fully combination reaction. The obtained samples were dark grey in color and stable in air for months.

The phase identification and crystal structure of the samples were characterized by powder X-ray diffraction (XRD) method on a PAN-analytical X-ray diffractometer with Cu-K$_\alpha$ radiation. The resistivity was measured by the standard four-probe method using a Quantum Design physical property measurement system (PPMS). The DC magnetization measurement was carried out with both field cooling (FC) and zero field cooling (ZFC) methods under a magnetic field of 5 Oe using a Quantum Design magnetic property measurement system (MPMS).

The powder XRD pattern for the polycrystalline LaRu$_2$As$_2$ sample at room temperature is presented in Fig. 1(c). All the diffraction peaks can be well indexed with the prototype of ThCr$_2$Si$_2$-structure with the space group of *I4/mmm* (No. 139), which is the same as the previously reported result [2]. The crystal structure was refined by the least-square fit method, which gave the lattice parameters of $a$ = 4.182(6) Å and $c$ = 10.590(3) Å. The XRD pattern indicates a single phase of the ThCr$_2$Si$_2$-type LaRu$_2$As$_2$ compound, and no other impurity phase was detected by XRD analysis in the sample. The schematic crystal structure of LaRu$_2$As$_2$ is also shown in Fig. 1(a), in which the Ru$_2$As$_2$ layers are separated by the La atoms. An image by scanning electron microscope in Fig. 1(b) shows the crystal grain morphology and the layered structure.

The temperature dependence of electrical resistivity for the $LaRu_2As_2$ sample was measured from 1.8 K to 300 K, and the data are shown in Fig. 2(a). As the temperature decreases, a sharp superconducting transition was observed. The onset superconducting critical temperature $T_c$(onset) is 7.8 K, and zero resistivity $T_c$(zero) happens at 6.8 K, with a transition width of 1 K for the polycrystalline sample. This $T_c$ observed in $LaRu_2As_2$ is much higher than that of the isostructural $LaRu_2P_2$ superconductor [2], and even all other iron-free transition metal pnictides with the $ThCr_2Si_2$-type crystal structure [22-27]. Above $T_c$, the resistivity data display a typical metallic behavior. The resistive superconducting transition was also characterized under various magnetic fields from 0 T to 1.8 T, with the temperature range between 1.8 K and 10 K. These data are shown in Fig. 2(b). As the magnetic field increases, the superconducting transition temperature $T_c$ decreases almost linearly, and the superconducting state above 1.8 K is totally suppressed under a magnetic field of 1.8 T. The upper critical field $\mu_0 H_{c2}(T)$ determined by the middle-point of resistivity transition at various temperature is shown in the inset of Fig. 2(a). The upper critical field at zero temperature $\mu_0 H_{c2}(0)$ was estimated to be 1.6 T by a linear extrapolation, which is much higher than that of the $LaRu_2P_2$ superconductor [28].

In Fig. 3(a), the temperature dependent DC magnetic susceptibility was measured between 2 K and 300 K with both zero-field-cooling (ZFC) and field-cooling (FC) methods under a magnetic field of 5 Oe. The superconducting Meissner effect happens at the onset temperature of 7.0 K in both ZFC and FC curves. The superconducting shielding volume fraction was estimated to be about 80% from ZFC data at 2 K, which indicates that the $LaRu_2As_2$ compound is a bulk superconductor. Above the superconducting transition, no other magnetic order was observed in the sample. The isothermal magnetization at 2 K was also measured, and the magnetic hysteresis exhibits that $LaRu_2As_2$ is a typical type-II superconductor.

In conclusion, we found superconductivity with a $T_c$ of 7.8 K in the ternary $LaRu_2As_2$ compound, and the magnetization measurements indicate this is a bulk type-II superconductor. Remarkably, the $T_c$ in LaRu$_2$As$_2$ is the highest value among the iron-free transition metal pnictides with the ThCr$_2$Si$_2$-type crystal structure.

**Acknowledgments**

The authors are grateful for the financial supports from the National Natural Science Foundation of China (No. 11474339), the National Basic Research Program of China (973 Program, No. 2010CB923000 and 2011CBA00100) and the Strategic Priority Research Program of the Chinese Academy of Sciences (No. XDB07020100).

**Figure Captions:**

Figure 1: (a) Schematic crystal structure of the LaRu$_2$As$_2$ compound. (b) A scanning electron microscope image for the polycrystalline sample. (c) Powder XRD patterns with indexed diffraction peaks for LaRu$_2$As$_2$ sample.

Figure 2: (a) The temperature dependence of electrical resistivity for the LaRu$_2$As$_2$ polycrystalline sample. The inset shows the upper critical fields $\mu_0H_{c2}$, determined at the middle-point of the resistivity transition. (b) The superconducting resistivity transitions under variable magnetic fields.

Figure 3: (a) The temperature dependence of magnetic susceptibility for the LaRu$_2$As$_2$ sample. The inset shows the expanded curve at the superconducting transition. (b) The isothermal magnetization curve for LaRu$_2$As$_2$ at 2 K.

**Fig. 1.**

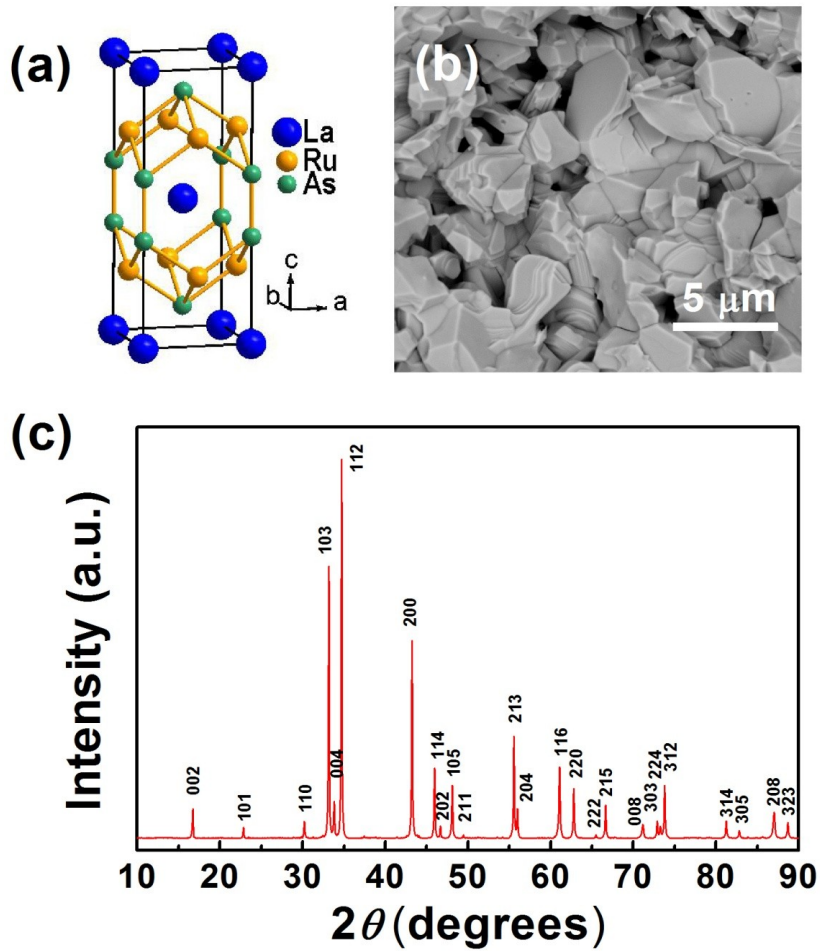

**Fig. 2.**

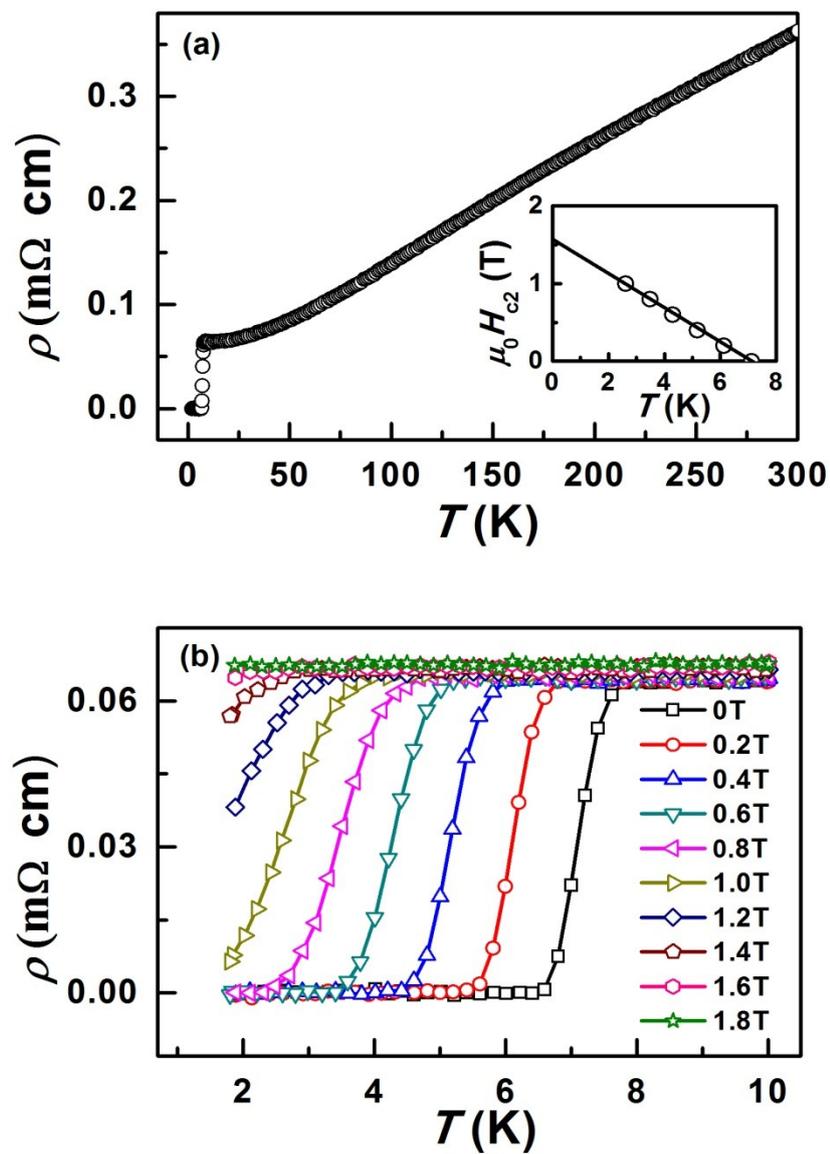

**Fig. 3.**

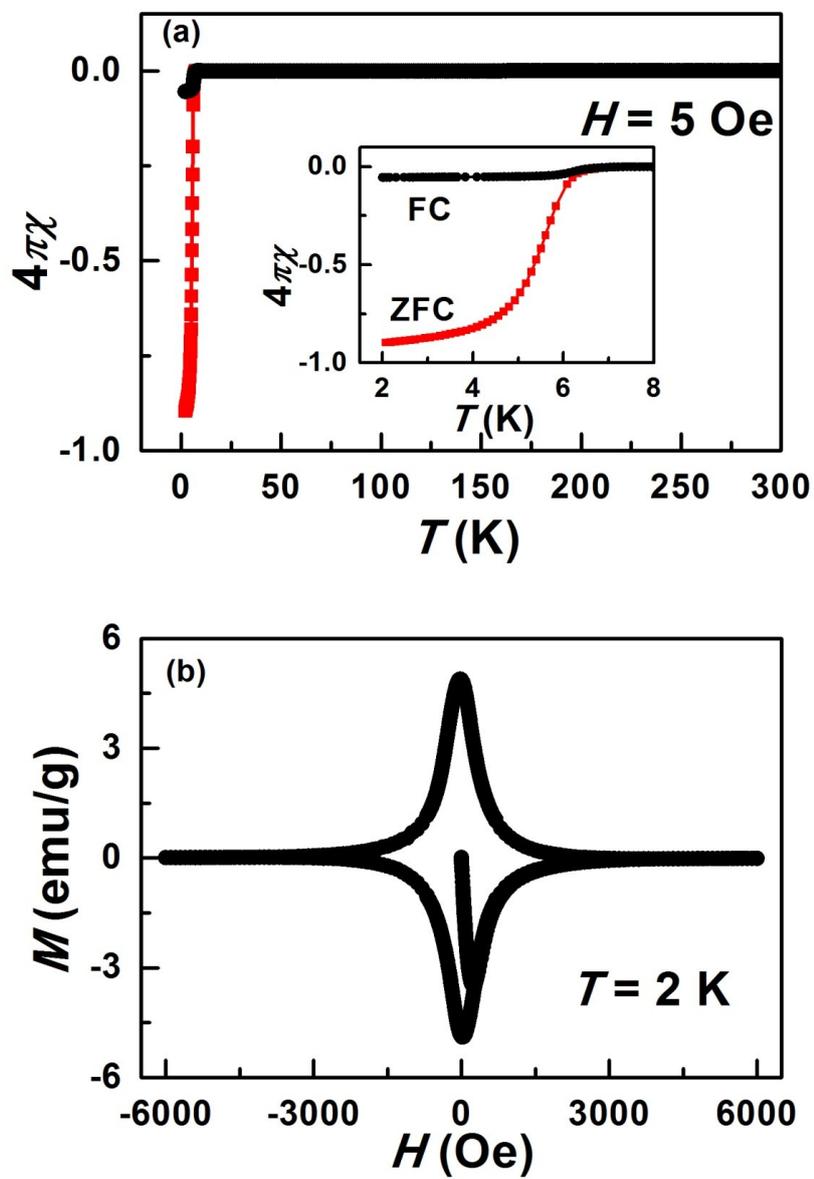